# Simulation of the growth kinetics in group IV compound semiconductors

*A. La Magna*, A. Alberti, E. Barbagiovanni, C. Bongiorno, M. Cascio, I. Deretzis, F. La Via, E. Smecca*

in memory of prof. Bengt Svensson

Consiglio Nazionale delle Ricerche Istituto per le Microelettronica e Microsistemi, Z.I. VIII Strada 5, I95121, Catania Italy
E-mail: antonino.lamagna@imm.cnr.it

Keywords: Kinetic Monte Carlo Simulations, Super Lattice Models, point defects, stacking faults, crystal growth

We present a stochastic simulation method designed to study at an atomic resolution the growth kinetics of compounds characterized by the $sp^3$-type bonding symmetry. Formalization and implementation details are discussed for the particular case of the 3C-SiC material. A key feature of our numerical tool is the ability to simulate the evolution of both point-like and extended defects, whereas atom kinetics depend critically on process-related parameters. In particular, the simulations can describe the surface state of the crystal and the generation/evolution of defects as a function of the initial substrate condition and the calibration of the simulation parameters. We demonstrate that quantitative predictions of the microstructural evolution of the studied systems can be readily compared with the structural characterization of actual processed samples.

## 1. Introduction

The crystal growth of group IV materials is a topic of extreme technological interest [1] due to the importance of these materials for current [2] and future technologies [3] (including quantum technologies, see e.g. [4]). In the case of compound semiconductors and in particular for SiC, the growth of a high quality material is particularly challenging due to the meta-stability of different crystal symmetries (polytypes) in the usual growth conditions [1]. The fundamentals on the growth kinetics are still not fully understood with accuracy and in general this lack of knowledge often results in a difficult process and material quality control in terms of defects and surface morphology. Analytic [5,7] theories are useful to categorize the phenomenology but cannot achieve the full predictive potential. Atomistic simulations could have access to the details of the growth mechanism that are beyond the possibility of experimental investigations, providing that: 1) a sufficiently accurate atom-atom interaction model is implemented in the simulation code; 2) the model is able to simulate kinetics at large time/space scales (growth time scales and mesoscopic crystal sizes).

Stochastic simulations in the Kinetic Monte Carlo (KMC) [8,9] approach could satisfy these two requirements, if the implemented model is correctly formulated for the class of systems under investigation and the calibration of the model parameters allows for the reproduction of the different growth regimes along with the related phenomenology. With this respect, Kinetic Monte Carlo formulations on augmented or super Lattices (KMCsL) [10-12] and parallel Lattices (KMCpL) [13] offer a great flexibility to simulate the structural evolution of defective systems [11,12] or the dynamical transition among different crystal structures [13].

This contribution reports the features of a stochastic KMCsL simulation code under development, with the aim of simulating at an atomistic resolution the growth process of compound materials characterized by $sp^3$ bond symmetry. The peculiar characteristic of this method is the possibility to study the generation and the evolution of extended and point defects during the growth process. In particular, in the class of extended defects, stacking faults (SF) are of primary interest due to their impact on the material quality and their negative consequences when building devices and applications. A specific focus will be given to the cubic SiC polytype (3C-SiC), but the approach is not limited to this particular compound semiconductor. We notice that previous KMC methods suffer from the one of the two following limitations: they either allow the defect simulation but do not simulate separately the evolution of the two system atoms (e.g. Si and C) in the compound (see Refs. [10,14]), or they simulate Si and C individually but the defect formation and evolution cannot be estimated (see Refs. [15,16]). In this paper we present a method satisfying both these requirements.

As we will explain in detail when discussing the formalism, the use of the super-Lattice formulation is a key element of the code in order to accommodate the defect configurations in the simulated systems. The manuscript sections are organized as follows: In Sec. 2, defect configurations are described comparing high-resolution transmission electron microscopy analyses with atomic models of the defects. In Sec. 3, the code formulation is presented and the implemented features are discussed. Results of KMCsL simulations are presented in Sec. 4 and in particular the ability of the code to simulate the generation and evolution of point and extended defects. Conclusive remarks and a plan of future advancements are reported in Sec. 5.

## 2. Atomistic Configuration of Stacking Faults in 3C-SiC

SFs are common and abundant extended defects in silicon carbide due to the similar energetics of different polytypes in such material. They can be easily categorized as wrong sequences with respect to the stacking order of the polytype in consideration. In the purely hexagonal close packed (hcp) representation the polytype order is the periodic sequence of layers, composed by Si-C dimers, along the hexagonal axis, where dimers in a single layer occupy one of the three highly symmetric positions (usually named A,B,C) of the hcp structure. The cubic 3C-SiC (zinc-blend) structure is obtained when the periodic sequence is ABC ABC ABC … .

In Fig. 1 the simplest case is shown, where only a single layer is not in the ideal crystal configuration. In the right panel of the figure, an image taken from a High Resolution Transmission Electron Microscopy (HRTEM) analysis performed with a sub-Angstrom JEOL ARM200F evidences the wrong atomic sequence corresponding to an intrinsic SF configuration for a 3C-SiC sample embedding the defect. The atomic reconstruction of the defect is shown in the left-hand panel where the faulted stacking sequence of the Si-C dimers along one of the <111> directions with respect to the ideal case is indicated as ABC ABC AC ABC. By observing the red lines in both figures, it can be easily understood that this defect causes a partial breaking of the epitaxial order in the two semi-spaces separated by the defect itself. Indeed, the atoms are displaced by 1/3 of the Si-C bond length along the bonds' direction. Another example of wrong stacking sequence is shown in Fig. 2, where three layers are not in the ideal crystal configuration. Similarly, the HRTEM analysis of the defective structure is shown in the right panel of Fig. 2. The wrong atomic sequence corresponds to a micro-Twin-like configuration. The atomic reconstruction of the defect is shown in the left-hand panel, where the faulted stacking sequence of the Si-C dimers along a <111> direction with respect to the ideal case is indicated as ABC ABC ACBA B. In this case, the micro-Twin defect does not break the epitaxial order in the two semi-spaces separated by the defect. The atoms in the two sides are perfectly aligned.

The atomistic mapping and the mechanism of the SF generation and evolution, which we would like to study with our KMCsL approach, must consider the local bonding configuration related to these defects. Indeed SF defects in 3C-SiC are due to a twist of the local bond network from the zig-zag like shape characterizing the zinc-blend crystal structure of the cubic SiC to the armchair one, typical of a wurzite like (i.e. hexagonal) order. In Fig. 3 the local bonding configuration of the regular (zig-zag) 3C-SiC crystal (grey and light brown balls) and the defective (arm-chair) types are shown: a $\pi/3$ twist of the three next neighbour atoms brings the regular configuration to the defective one, where the rotation axis coincides with the remaining next-neighbour bond.

**3. Kinetic Monte Carlo super-L Model**

According to the augmented lattice paradigm, our KMCsL model is based on a super-Lattice able to accommodate correctly the defect configurations, where the original lattice of the ideal crystal is a sub-lattice of the super lattice itself. As a result of (a) a series of HRTEM analyses (two of them already discussed in Sec. 2) and (b) the symmetry considerations on zinc-blend, wurzite and in general any hexagonal polytype of SiC, we have deduced that a reliable super-Lattice formulation of our problem can be only obtained using a cubic reference lattice with constant $a_{KMCsL}=l_{SiC}/(27)^{0.5}$, where $l_{SiC} \approx 0.185$nm is the Si-C bond length. We note that this super-Lattice is a relative dense lattice with respect to the original zinc-blend one (for example two nearest neighbors in the regular lattice are the sites (0,0,0) and (3,3,3) in units of $a_{KMCsL}$). In order to understand the relation between the regular and the super Lattice in the KMCsL description of Fig. 4, the zinc-blend 3C-SiC tetrahedral configuration is shown embedded in the portion of the super-Lattice sites which can be obtained by $\pi/3$ twists of the regular sites.

It is noteworthy that both regular and defective configurations can be locally reconstructed as sites of the super lattice in a way that is easy to implement in a computational algorithm dealing with a crystal growth (etch) problem. Indeed, looking at Fig. 3, if we assume that SiteC has already fixed its local coordination symmetry forming some bonds with its next neighbours (indicated by arrows in the figure) then: (a) the regular (zig-zag) next-neighbour sites of the "(New)Site" are the symmetric positions of the next-neighbours of "SiteC" with respect to the centre of the SiteC-(New)Site bond, while b) the arm-chair next neighbour sites of the "(New)Site" can instead be obtained by displacing the next-neighbours of "SiteC" by a length of $5\times a_{KMCsL}$ along the direction of the SiteC-(New)Site bond.

**3.1 KMCsL events and particles**

Our kinetic code assumes that the growth occurs in the surface layer between the gas phase and the crystal phase due to a balance between atomic transitions from/to one phase to the other. The following events are considered: a) attachments of Si or C atoms on a dangling bond present at the solid-gas interface ("*Deposition*" event in the following), b) detachments of not fully (four fold) coordinated atoms, which pass from the solid to the gas phase ("*Evaporation*" event in the following). Consistently, the evolution is driven by "active" MC particles and they are of two types: 1) under coordinated atoms (coordination from one to three) here and after named "Ad-atoms", 2) super-Lattice empty sites which are connected to an "Ad-atom" by one of its dangling bonds, here and after named "Ad-void". *Evaporation* is the MC event which transforms an "Ad-atom" in "Ad-void" while *Deposition* is the MC event which promotes the "Ad-void" to "Ad-atom". In fig. 5 some examples of *Deposition/Evaporation* events are reported and the local reconfigurations due to the Ad-void/Ad-atom transformations are shown for these particular cases.

In the KMC implementation, the stochastic event selection is only ruled by probability rules and all the MC particles are considered at the same level. As a consequence, Ad-voids and Ad-Atoms are stored in the same list of MC particles here and after called "*ListAdAtom*" (i.e. "solid" and "gas" atoms). In addition to this list, we also store two lists of "silent" particles i.e. the 4-fold coordinated atom in the Bulk crystal (*ListAtom*) and the 4-fold coordinated vacancy sites (*ListVoid*). These "silent" particles can become MC particles during the evolution in the case when one of their four next neighbor Ad-atoms undergoes an *Evaporation* event.

The list stores the lattice locations of the particle, the positions of their next-neighbor sites and in the case of the MC particles also the type of event and the related probability. For each particle there is a "bit encoded" amount information stored in the integer array *LattCoo*, which stores the super lattice locations (x,y,z), namely: the index in the *List*, the occupancy state, the coordination value (*Coor*) and the type of particle.

**3.2 Zig-Zag/Arm-chair bonds selection**

As already stated, the distinguished feature of our KMCsL code is the possibility to consider defect generation and evolution whilst the *$sp^3$ like hybridization of the atomic orbitals is assumed as fixed* (we note that this is a general constrain of our method). We note that, due to this $sp^3$ bonding constrain, for all *Ad-Atoms* or *Ad-Voids* with *Coor*>1, the remaining bond directions are fixed. When *Coor*=1, the local symmetry has in principle two possible degenerate configurations (i.e. the zig-zag and the armchair one). We introduce the MC rule to fix stochastically the local bond configurations in the *Deposition* event for an *Ad-atom* with *Coor*=1 (i.e. when the KMC algorithm selects an *Ad-void* with *Coor*=1 as the current evolving particle for an *Ad-void* to *Ad-atom* transition). As a consequence, after this event the *Ad-Atom* will chose one of the two possible configurations according to the algorithm presented in the following. Preliminarily, from these previous considerations we deduce that:

- All the Ad-Atoms and all the Bulk Atoms and Voids have fixed dangling-bond directions;
- Ad-Voids with *Coor*=2,3 have fixed dangling bond directions;
- Ad-Voids with *Coor*=1 do not have fixed dangling bond directions.

The stochastic selection rule of the coordination type considers: a) the epitaxial relationship between the new Ad-atom and the surrounding crystal Atoms/Ad-atoms, and b) the probability of defective configurations generation. Therefore the selection algorithm proceeds as follows (see again Fig.3):

If New Site is an Ad-Void selected for a deposition transition with *Coor*=1 and SiteC is its (single) next-neighbor Atom or Ad-Atom, (note that isolated atoms are instantaneously removed by the simulation, therefore the next-neighbors of SiteC are fixed); the local Zig-Zag and Arm-Chair bonding configurations are determined according to the following algorithm:

- we define *CoordSumZig* as the sum of the coordination of all "Zig-Zag" Sites;
- we define *CoordSumArm* as the sum of the coordination of all "Arm-Chair" Sites;
- if *CoordSumZig* > *CoordSumArm* (*CoordSumZig* < *CoordSumArm*) the Zig-Zag (Arm-Chair) configuration is chosen;
- *if CoordSumZig* = *CoordSumArm*, the configuration is stochastically chosen using the calibration parameter *PtransZigZag*: if rand(0,1) < *PtransZigZag* (rand(0,1) > *PtransZigZag*) the *Zig-Zag* (*Arm-Chair*) case is picked up (rand(x,y) is a random number in the open interval indicated).

Note that *if PtransZig=1 no defect forms in a 3C-SiC growth simulation.* Finally, a natural prescription of reciprocity is imposed in the coordination evaluation: *a bond can form only if two next neighbour MC particles point the relative orbitals in the same direction*.

**3.3 Deposition**

*Deposition* causes the promotion of one *Ad-void* in the position *Site* to *Ad-atom* (it remains an MC particle but the event associated to this particle will be the opposite one i.e. *Evaporation*). After the *Deposition,* the Site coordination does not change. The associated *Coor*(NN) of the Next Neighbour (NN) Sites increases by 1 (note that after the event all the Next Neighbours of Site will be surely MC particles: Ad-Atoms or Ad-Voids). As discussed, if Coor(*Site*) = 1, the associated dangling bond direction will be chosen (Zig-Zag or Arm-Chair with proper probability rule).

Apart from the change in the evolving particle (Site) state (see also fig. 5, where reconfigurations due to the *Deposition* in Ad-Void sites with Coor=1 and Coor=2 is shown), due to the *Deposition* event the information stored in the *Lists* and in the super-lattice (*LattCoo*) must be updated, depending on the type and coordination of the neighbors after the event, as in the following:

- Case *Void* and *Coor*(NN)=1. This site will become an MC particle with no fixed dangling bond direction.

- Case *Void* and *Coor*(NN)=2. This MC particle will fix its dangling bond direction (Site + its previous NN).
- Case *Void* and *Coor*(NN)=3. This is an MC particle with the dangling bond direction already fixed.
- Case *Void* and *Coor*(NN)=4. This site will not be an MC particle after the Deposition: a Vacancy site is generated (i.e. it is removed from the list of the MC sites (*ListadAtom*) and it is added to the list of the Vacancy sites (ListVoid).
- Case *Atom* and *Coor*(NN)=2-3. This MC particle has already a fixed dangling bond direction.
- Case *Atom* and *Coor*(NN)=4. This site becomes a Bulk particle with fixed dangling bond direction (i.e. it is removed from the list of the MC site (*ListAdAtom*) and it is added to the list of the Bulk sites (*ListAtom*)).

**3.4 Evaporation**

*Evaporation* causes the transition of one Ad-atom to an Ad-void in the position Site. The Site remains related to an MC particle but the event associated to this particle will be the opposite one i.e. the *Deposition*. Similarly to the *Deposition,* after *Evaporation* the Site's coordination does not change. The associated *Coor*(NN) of the NN sites decreases by 1 (for all the Next Neighbour MC particles Atoms and Voids, which in this case can be eventually erased by the ListAdAtom if *Corr*(NN)=0). If *Coor(Site) = 1 the associated dangling bond direction will be lost* and the Next Neighbour list of the ListAtAtom associated variable will contain only the single Atom type site among the Next Neighbours.

Apart from the change in the evolving particle (Site) state, due do the *Evaporation* events the information stored in the Lists and in the super-lattice (*LattCoo*) must be updated, depending on the type and coordination (after the *Evaporation*) of the neighbors, as in the following (see fig. 5 for some particular examples):

- Case *Void* and *Coor*(NN)=0. This is not an MC particle anymore; the associated information will be deleted from lists and lattice.
- Case *Void* and *Coor*(NN)=1. This MC particle (*Coor* passes from 2 to 1) will loose the information related to the dangling bond directions.
- Case *Void* and *Coor*(NN)=2. This MC particle will maintain its state, with the dangling bond direction fixed.
- Case *Void* and *Coor*(NN)=3. This site was a vacancy before the *Evaporation*. The vacancy site will become an active MC particle with the dangling bond direction fixed.
- Case *Atom* and *Coor*(NN)=1-2. This MC particle will maintain its state with already fixed dangling bond direction.
- Case *Atom* and *Coor*(NN)=3. This was a bulk fully coordinated atom before the *Evaporation*. It will become an MC particle with fixed dangling bond direction (i.e. it is removed from the list of the Bulk sites (ListAtom) and it is added to the list of the MC sites (ListAdAtom).

### 3.5 Event Selection and Probability Table

The evolution in the Kinetic Monte Carlo method is simulated by means of a sequence of events which are picked up according to the associated probability (frequency) of each event, where an event with larger probability has a greater chance to be selected with respect to one with a lower probability. However, also rare events can be selected during the simulated kinetics. In our code, we apply continuous time and tree selection methods, which we briefly outline in this section. The n-th step of the simulation has an associated probability/frequency table

$$\{\omega_n(i)\} \tag{1}$$

where the index $i$ runs on all the possible events (all the *Deposition* and *Evaporation* in our case). The time increment $\Delta t_n$ at the n-th step is calculated by means of the total probability using the following equations

$$\Omega_n = \sum_i \omega_n(i) \Rightarrow \Delta t_n = \frac{1}{\Omega_n} = \frac{1}{\sum_i \omega_n(i)} \tag{2}$$

Then the "occurring" event should be selected according to its conditional probability:

$$P_n(i) = \frac{\omega_n(i)}{\sum_i \omega_n(i)} \tag{3}$$

This result could be statistically obtained with a chain selection rule

$$\omega_n(\bar{i}) \le rand_n(0,1) \times \left[ \sum_i \omega_n(i) \right] < \omega_n(\bar{i}+1) \tag{4}$$

whose computational cost grows as the number N of total events. This number is usually very large (e.g. N>>$10^5$) and as a consequence we decided to adopt the equivalent (stochastically) binary tree selection method [5]. The cost of this method is $o(\log_2(N))$ and it overpasses in the balance the cost of the tree management and update.

The KMCsL model's parameters used to determine the set $\{\omega_n(i)\}$ are shown in the tables 1-3. They depend on the total local coordination and type (Si or C) of the Ad-Void for the *Deposition* event; whilst for the *Evaporation* event, when fixing the total coordination number (1,2,3 or 4) of the Ad-Atom, different probabilities/frequencies of events are assigned for different compositions of next neighbor atoms (where for *Coor*(NN)=4 the standard choice for the evaporation probability is zero in order to avoid the evaporation of bulk atoms). This option allows increasing the stability of the ideal alternation of neighbors (i.e. the Si (C) atom has a C (Si) next neighbor in SiC).

**4 Results**

In this section, we will discuss some results of demo simulations performed with our code. The scope is not to simulate real growth processes but only to show the possible features of the code. Indeed, a realistic simulation can be obtained after a reliable calibration of the parameters shown in Tab.1. We note that these in general should depend also on the macroscopic process parameters as the local-temperature, partial pressures of the gas components, etc. Future work will be dedicated to derive the calibration setting (i.e. the correlation between macroscopic process parameters and the KMCsL atomistic event frequencies of tables 1-3) for real process cases. Anyhow, the scope of this section is to demonstrate which information could be extracted by the simulation, once the calibration has been validated. In this section, we fix the *Evaporation* parameters as indicated in the Tabs 4 and 5 while we vary the *Deposition* parameters and the value of *PtransZigZag.*

The particular choice of the evaporation parameter will favor the Si-C type bond network, since a fast evaporation should occur in the case of formation of homo-atomic bonds (C-C or Si-Si type). Moreover, the configurations with a larger coordination are more stable. The symbol $\omega$ indicates an average frequency which sets the time scale.

The code simulates the evolution as a balance of *Evaporation/Deposition* events. As a consequence, the simulation will reproduce deposition or etching according to the relative values of the related frequencies. To demonstrate this feature in Fig. 6 we show two simulation snapshots obtained for different choices of the deposition parameters. The relative height with respect to the initial substrate is positive (negative) in the case of the top (bottom) panel; as a consequence, both deposition and etching can be simulated. Please note that in order to facilitate the visualization in this and in the following figures, only under-coordinated atoms are shown. In the deposition case (top-panel) we can also observe the generation of point defects of the vacancy type: the empty site is indicated by a red point and its under-coordinated next-neighbors are also visible.

Vacancy generation efficiency could be experimentally investigated in post-process characterization analyses and it could be a useful topic for the code use as a support of the experimental results. We could expect that vacancy generation depends on the process macroscopic conditions, as this aspect can be considered in the code parameter calibration. Indeed, the density of vacancies trapped in the growing crystal in a simulated growth depends strongly on the parameter choice. In Fig. 7 snapshots of simulated growths are shown, where the dependency of vacancy generation on the deposition parameters is evident. Please note that in the demo simulation here presented we use a symmetric calibration for the two atomic species (Si and C); as a consequence similar (i.e. equivalent, due to the stochastic approach) densities if C-type and Si-type vacancies are generated. Of course, eventual asymmetric generation of one type of vacancy, which could emerge from experimental evidences, can be considered with a proper calibration choice.

Evolutions of nanostructures and in general of non-planar substrates will be one of the main application fields of the developed KMCsL code. In Fig. 8 the growth of a nano-crystal starting from a cubic seed is shown for a calibration setting equal to the one indicated in the top panel of Fig. 6. A similar shape evolution is obtained (not shown) using a spherical seed. The facets of the nano-grains are aligned along the <100> and <111> directions and the structure is rather regular. Please note that the nano-crystal shape also depends on the calibration choice. For example, using a symmetric parameter evaporation setting we can effectively simulate the case of a mono-atomic crystal growth (e.g. a pure Si grain) and in this case (not-shown) the crystal shape is similar to an octahedron with dominant <111> facets and <311> type edges.

Analyses of SiC nano-crystal morphology and microstructures generally show both a rather irregular nano-crystal shape and the occurrence of a strongly defective state dominated by stacking-fault-type extended defects. Our KMCsL code reproduces this behavior when a value PtransZigZag$<1.0$ is used in the calibration and the code simulates possible local switches to armchair like bonds. In Figs. 9 and 10 we report the results of two simulations of nano-crystal growths using PtransZigZag=*0.8* and PtransZigZag=*0.5* respectively, while the rest of the parameters are fixed as in the case of Fig. 6 top panel.

In the bottom panels of Figs. 9 and 10 we show only the atoms which are not in epitaxial order with respect to the initial seed. In both cases, we observe the formation of stacking faults in the <111> facets of the growing nano-crystal. For the larger value of PtransZigZag=*0.8*, the defect formation is less probable and the simulation shows the presence of some single plane wrong stacking: i.e. the intrinsic SF shown in Fig. 1. Of course, as discussed in Section 2, all atoms in one side of the defect are not in epitaxial position with respect to the seed. In the case of PtransZigZag=*0.5* the switch probability is larger and the generation of extended defects is strong. SFs form in all the 8 <111> facets in this simulated growth. Moreover, the larger probability of consecutive switches causes the formation of different symmetries of SFs where a predominance of the micro-Twin like (Fig. 2) is computed in the simulated growth.

We note that also the global morphology of the nano-crystal is affected by the defect formation and more irregular shapes are simulated with respect to the case of PtransZigZag=*1.0* (Fig. 8).

**5 Conclusion**

In this manuscript we discussed the formulation, implementation and capabilities of a KMCsL code able to simulate the growth of compound semiconductors with $sp^3$-type bonding. The 3C-SiC case is considered as a reference system for the simulation. The results here shown clearly demonstrate some code features which are appealing for its application as a support of experimental activity related to material growth and characterization. Here we summarize the main ones:

- The code has the possibility to use any surface (polar/non-polar/structured), whereas some previous approaches fixed an exposed polar surface and its hexagonal stacking;

- The atomic species of the compound (here Si and C) can undergo independent kinetics;
- The code allows for the study of point and extended defect generation at an atomic level, considering a non-ideal Si-C bond configuration;
- The coupling of the atomistic code with gas phase simulations (or experimental gas kinetic studies) is possible due to the independent kinetics of Si-type and C-type reactions in order to assess the code calibration;

Future activity will be dedicated to the cited parameter calibration and code validation by means of comparisons with available process data. A preliminary and partial calibration can be obtained using Arrhenius-type expressions from the event frequencies in tables (1-3) and using bonding energies available in the literature (see e.g. Ref [17,18]). However, not all the model parameters have a counterpart in literature data and additional parameters' evaluations are required also with the aid of ab-initio calculations [12]. Moreover, the strategy to achieve a reliable and accurate calibration assessment could also rely on the code features starting from an initial calibration set. The possibility to estimate several "observables" (e.g. the speed of growth, the point defect density, the SF configuration and evolution, the shape evolution of microcrystalline objects, etc.) allows for an effective experimental verification of the code predictions, since simulation "observables" are either measurable quantities or microstructures accessible to the characterization by means of advanced techniques. As for any parameter-reliant approach, a refinement of the calibration will be therefore necessary by fine tuning the parameters in order to link the simulation predictions to the experimental data for variable thermodynamic conditions of the growth processes.

**Acknowledgements**

This work has been partially supported by the CHALLENGE project (HORIZON 2020 – NMBP – 720827, http://www.h2020challenge.eu/), CHALLENGE is a research and innovation action funded by the European Union's Horizon 2020 programme.

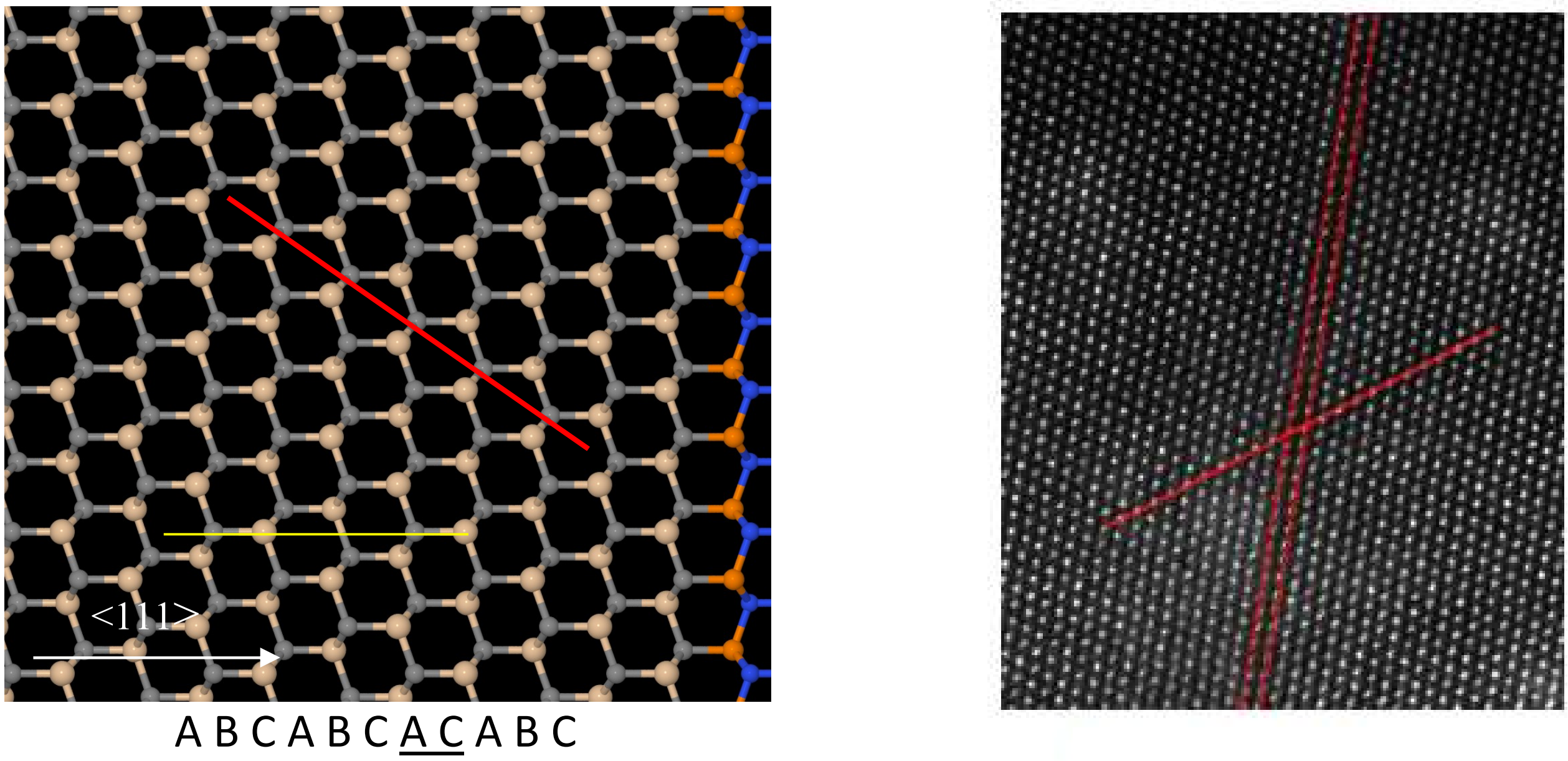

**Fig. 1** Atomic representation (Left-hand panel) and HRTEM (right-hand panel) section of an "intrinsic" SF in 3C-SiC. The faulted stacking sequence of the Si-C dimers along a <111> direction with respect to the ideal ABC ABC ABC…one is indicated. The red line is a guide for the eye, highlighting the breaking of the epitaxial order in the two semi- spaces separated by the defect: The atoms are displaced by 1/3 of the Si-C bond distance.

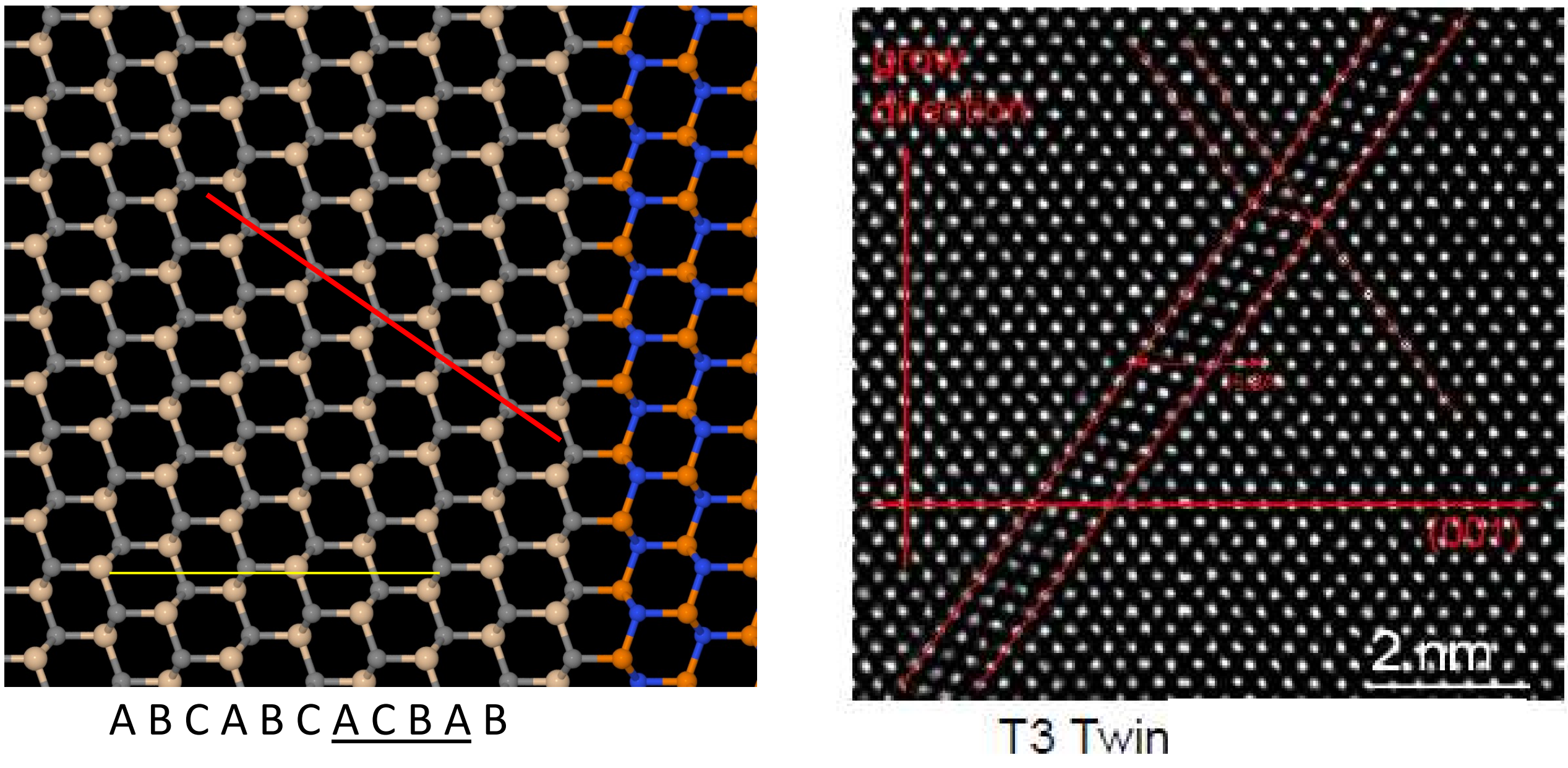

**Fig. 2** Atomic representation (Left-hand panel) and HRTEM (right-hand panel) image of a "Micro Twin like" SF in 3C-SiC. The faulted stacking sequence of the Si-C dimers along a <111> direction with respect to the ideal ABC ABC ABC…one is indicated. The red line is a guide for the eye, highlighting the maintained epitaxial order in the two semi-spaces separated by the defect.

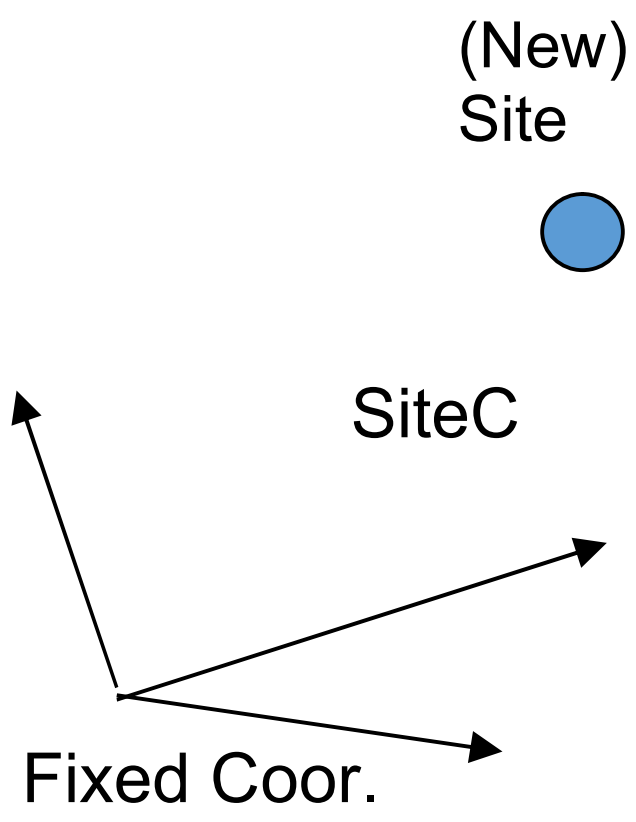

**Fig. 3** Local regular zig-zag (grey and light brown spheres) and defective armchair (yellow spheres) configurations for a zinc-blend structure. In the super-Lattice description, the regular next neighbour sites of the "(New)Site" are the symmetric positions of the next-neighbours of "SiteC" with respect to the centre (blue circle in the right-hand panel) of the SiteC-(New)Site bond. The defective next neighbour sites of the "(New)Site" can be instead obtained displacing the next-neighbours of "SiteC" by a length of $5\times a_{KMCsL}$ along the direction of the SiteC-(New)Site bond.

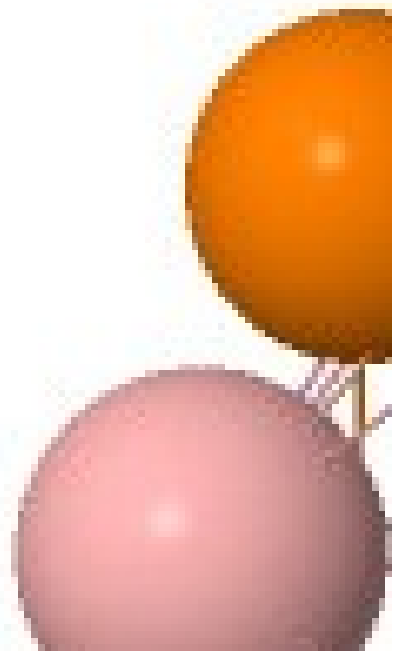

**Fig. 4** Zinc-blend tetrahedral configuration (grey and light brown spheres) as filled site of a sub-lattice embedded in our super lattice. Other sites of the super lattice that can be obtained by π/3 twists of the regular sites are shown as pink and orange spheres.

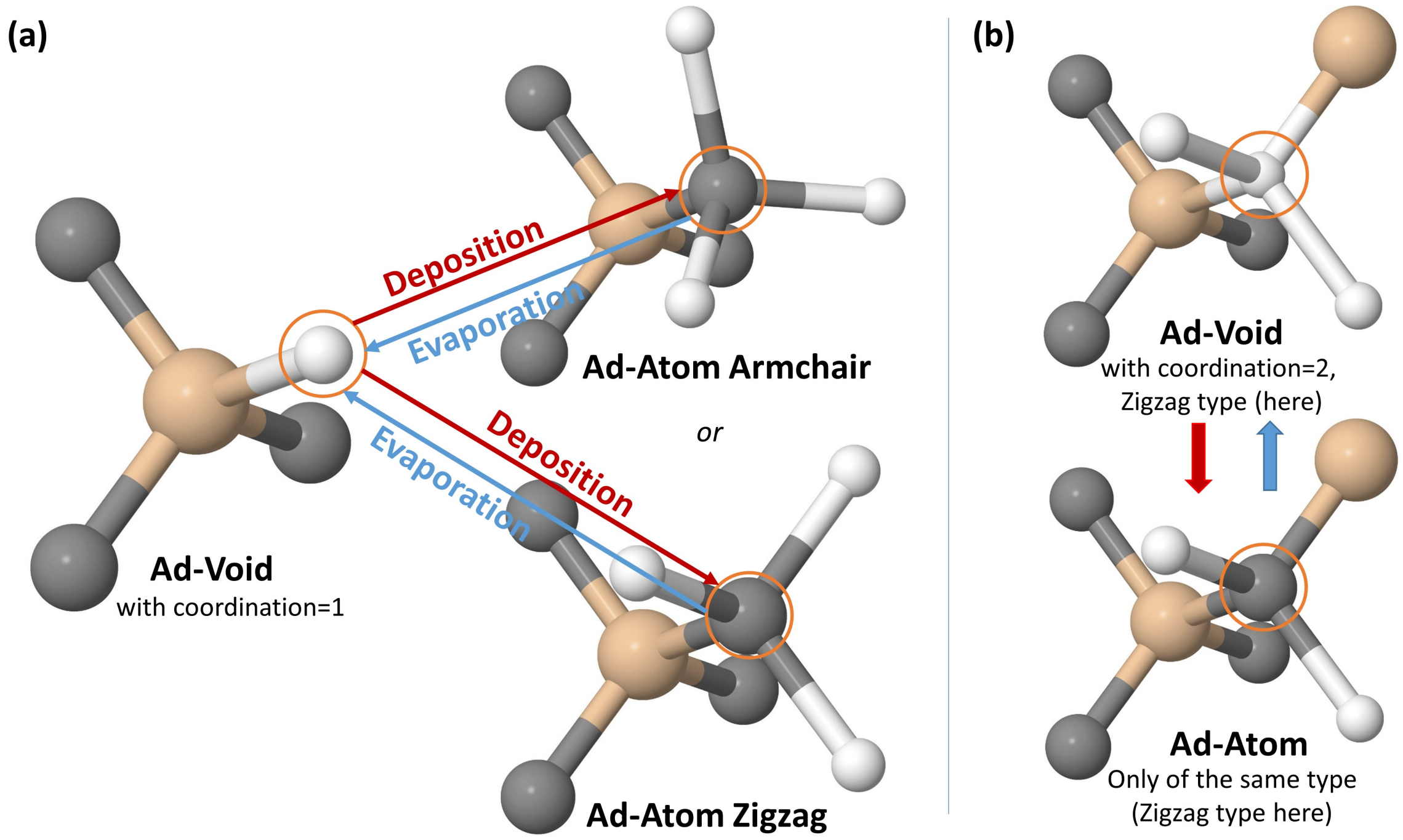

**Fig. 5** Schematic of the local bonding rearrangement due to deposition/evaporation events in the case of coordination value *Coor*=1 [panel (a)] and Coor=2 [panel (b)]. Large grey and light brown spheres indicate *AdAtoms* (C and Si) while small white spheres indicate *AdVoids*. For a deposition event in the case of *Coor*=1 the local bonding configuration (Armchair o Zigzag) has to be statistically selected. The bonding network is undetermined for an AdVoid with *Coor*=1 while it is fixed for both and *AdVoid* and *AdAtom* with Coor=2, see e.g. panel (b) where a Zigzag type of bonding is shown.

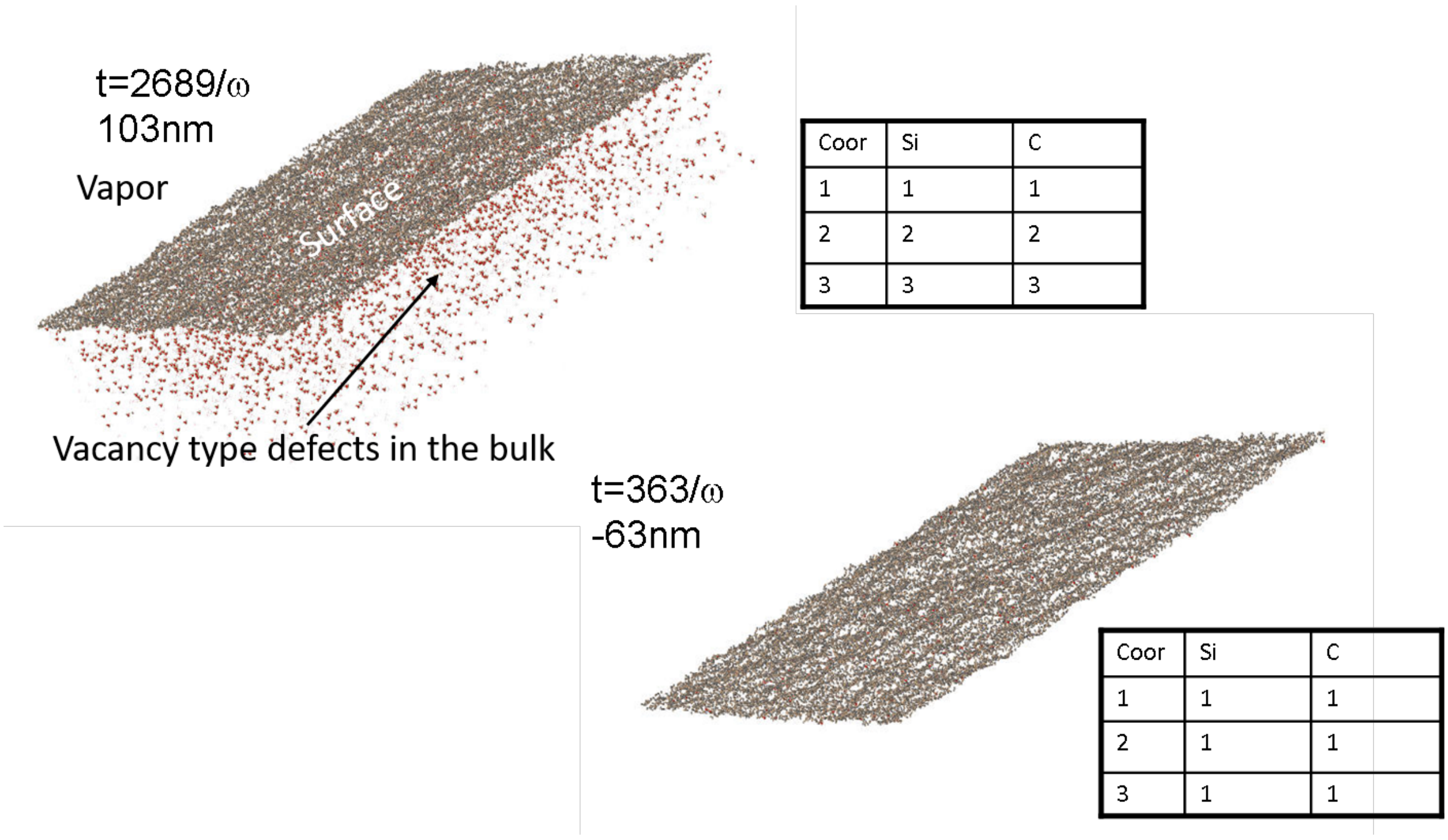

| Coor | Si | C |
|---|---|---|
| 1 | 1 | 1 |
| 2 | 2 | 2 |
| 3 | 3 | 3 |

| Coor | Si | C |
|---|---|---|
| 1 | 1 | 1 |
| 2 | 1 | 1 |
| 3 | 1 | 1 |

**Fig. 6** Simulation snapshots obtained starting from a flat 3C-SiC substrate for two different choices of the deposition parameters (in ω units) and PtransZigZag=1. The height relative to the initial substrate obtained in the two simulated evolutions is indicated. In the top (bottom) panel a growth (etching) process is simulated. Vacancy type defects are visible in the growth case. They are in the bulk crystal whist the vapor phase is in the other side with respect to the surface (these regions are indicated in the top panel).

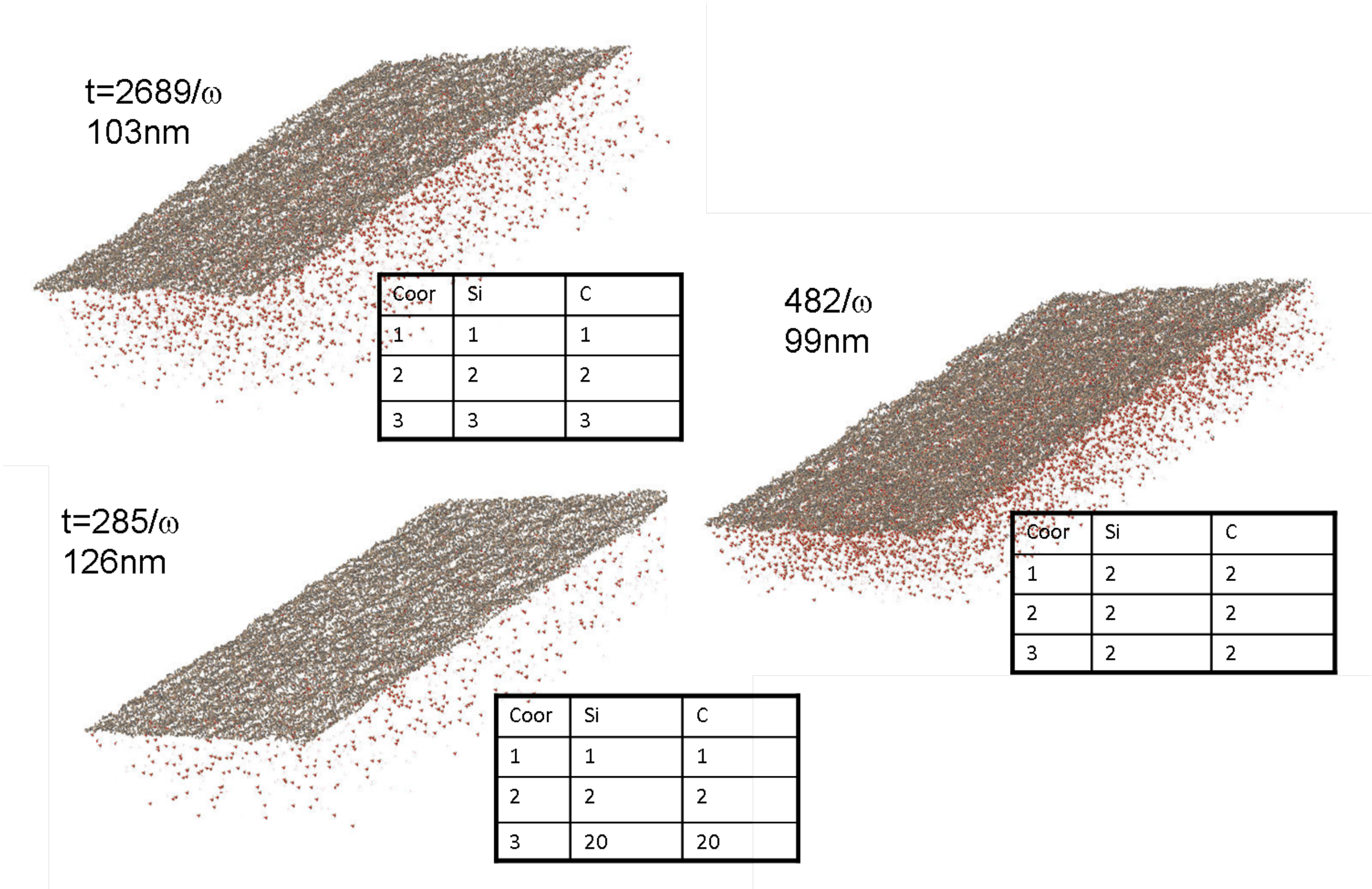

**Fig. 7** Simulation snapshots obtained starting from a flat 3C-SiC substrate for three different choices of the deposition parameters (in ω units) and PtransZigZag=1. The quote relative to the initial substrate height obtained in the simulated evolution is indicated. Vacancy type defects are visible in all the graphs. They are in the bulk crystal whist the vapor phase is in the other side with respect to the surface.

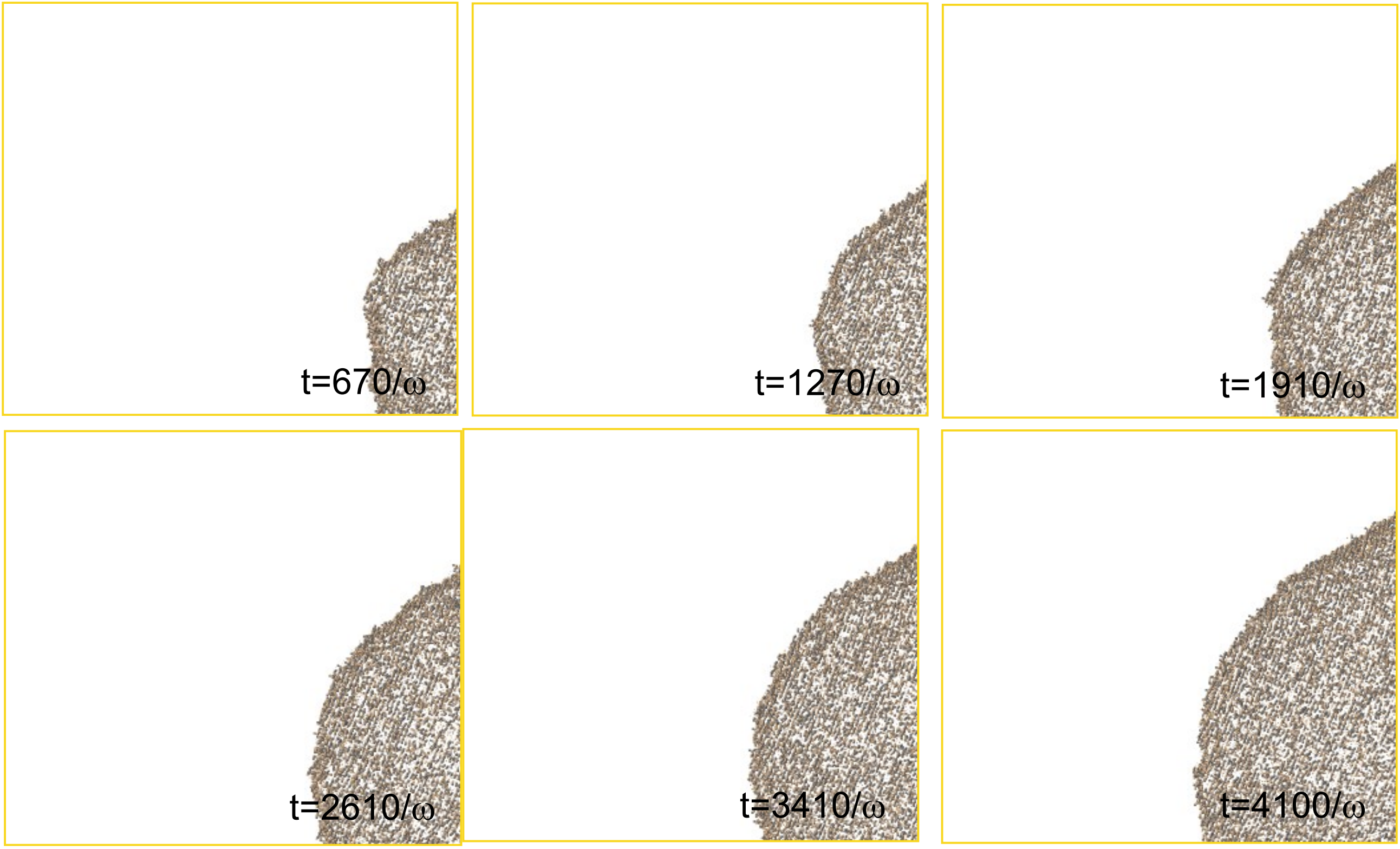

**Fig. 8** Simulation snapshots of the growth of a 3C-SiC nano-crystal starting from a cubic seed with facets oriented in the <100> direction. The deposition parameters are equal to the case reported in Fig. 5 top panel.

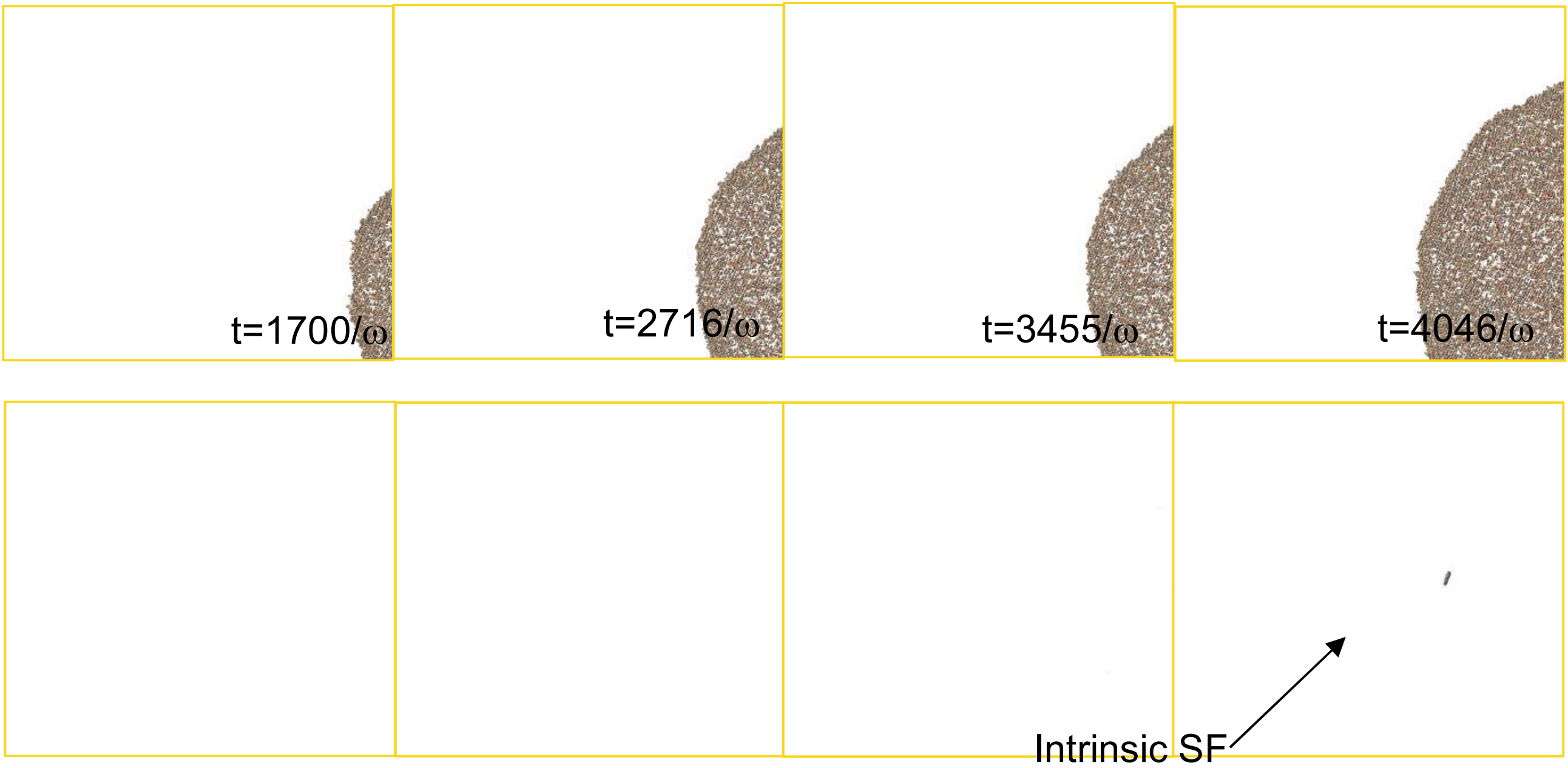

**Fig. 9** Simulation snapshots of the growth of a 3C-SiC nano-crystal starting from a cubic seed with facets oriented in the <100> direction. In the bottom panel, atoms which are not epitaxially ordered are shown. The deposition parameters (in ω units) are equal to the case reported in Fig. 5 top panel (PtransZigZag=0.8).

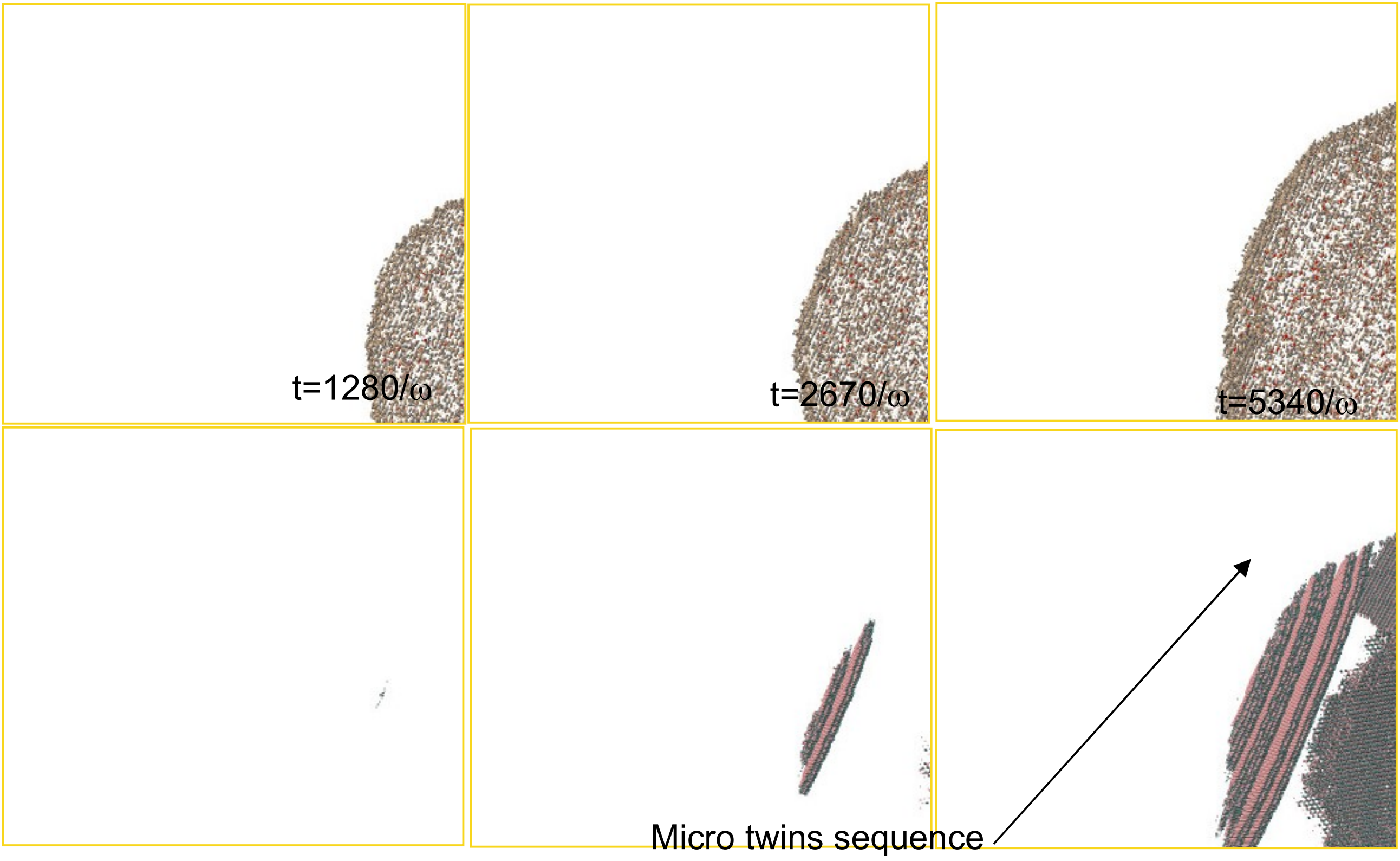

**Fig. 10** Simulation snapshots of the growth of a 3C-SiC nano-crystal starting from a cubic seed with facets oriented in the <100> direction. In the bottom panel, atoms which are not epitaxially ordered are shown. The deposition parameters (in ω units) are equal to the case reported in fig. 7 top panel (PtransZigZag=0.5).

***Tab. 1.*** Calibration parameters of the *Deposition* events

| Coordination | Atom_1 (e.g. Si) | Atom_2 (e.g. C) |
|---|---|---|
| 1 | $P_{Dep}(1,1)$ | $P_{Dep}(2,1)$ |
| 2 | $P_{Dep}(1,2)$ | $P_{Dep}(2,2)$ |
| 3 | $P_{Dep}(1,3)$ | $P_{Dep}(2,3)$ |

***Tab. 2*** Calibration parameters of the *Evaporation* events for Atom_1 (e.g. Si)

| NN C or Si | NN Atom_1=0 | NN Atom_1=1 | NN Atom_1=2 | NN Atom_1=2 |
|---|---|---|---|---|
| NN Atom_2=0 | - | $P_{Evap_1}(1,0)$ | $P_{Evap_1}(2,0)$ | $P_{Evap_1}(3,0)$ |
| NN Atom_2=1 | $P_{Evap_1}(0,1)$ | $P_{Evap_1}(1,1)$ | $P_{Evap_1}(2,1)$ | $P_{Evap_1}(3,1)$ |
| NN Atom_2=2 | $P_{Evap_1}(0,2)$ | $P_{Evap_1}(1,2)$ | $P_{Evap_1}(2,2)$ | - |
| NN Atom_2=3 | $P_{Evap_1}(0,3)$ | $P_{Evap_1}(1,3)$ | - | - |

***Tab. 3*** Calibration parameters of the *Evaporation* events for Atom_2 (e.g. C)

| | NN Atom_1=0 | NN Atom_1=1 | NN Atom_1=2 | NN Atom_1=3 |
|---|---|---|---|---|
| NN Atom_2=0 | - | $P_{Evap_2}(1,0)$ | $P_{Evap_2}(2,0)$ | $P_{Evap_2}(3,0)$ |
| NN Atom_2=1 | $P_{Evap_2}(0,1)$ | $P_{Evap_2}(1,1)$ | $P_{Evap_2}(2,1)$ | $P_{Evap_2}(3,1)$ |
| NN Atom_2=2 | $P_{Evap_2}(0,2)$ | $P_{Evap_2}(1,2)$ | $P_{Evap_2}(2,2)$ | - |
| NN Atom_2=3 | $P_{Evap_2}(0,3)$ | $P_{Evap_2}(1,3)$ | - | - |

***Tab. 4.*** Parameters of the *Evaporation* events for Atom_1 (e.g. Si) used in the demo simulations discussed

| | NN Atom_1=0 | NN Atom_1=1 | NN Atom_1=2 | NN Atom_1=3 |
|---|---|---|---|---|
| NN Atom_2=0 | - | 50.0ω | 20.0ω | 10.0ω |
| NN Atom_2=1 | 3.0ω | $P_{Evap_1}(1,1)$ | $P_{Evap_1}(2,1)$ | 0.0 |
| NN Atom_2=2 | 2.0ω | $P_{Evap_1}(1,2)$ | 0.0 | - |
| NN Atom_1=3 | 0.1ω | 0.0 | - | - |

**Tab.5** Parameters of the *Evaporation* events for Atom_2 (e.g. C) used in the demo simulations discussed

| | NN Atom_1=0 | NN Atom_1=1 | NN Atom_1=2 | NN Atom_1=3 |
|---|---|---|---|---|
| NN Atom_2=0 | - | 3.0ω | 2.0ω | 0.1ω |
| NN Atom_2=1 | 50.0ω | 20.0ω | 10.0ω | 0.0 |
| NN Atom_2=2 | 20.0ω | 10.0ω | 0.0 | - |
| NN Atom_2=3 | 10.0ω | 0.0 | - | - |